\documentclass[sigconf]{acmart}

\usepackage{algorithm}
\usepackage{algpseudocode}
\usepackage{booktabs}
\usepackage{multirow}
\usepackage{enumitem}
\usepackage{bm}
\usepackage{graphicx}
\usepackage[normalem]{ulem}
\usepackage{xspace}
\usepackage{adjustbox}
\usepackage{marvosym}
\usepackage{balance}
\usepackage{float}

\usepackage{bbding}

\AtBeginDocument{%
  \providecommand\BibTeX{{%
    \normalfont B\kern-0.5em{\scshape i\kern-0.25em b}\kern-0.8em\TeX}}}

\copyrightyear{2022}
\acmYear{2022}
\setcopyright{acmcopyright}
\acmConference[KDD '22] {Proceedings of the 28th ACM SIGKDD Conference on Knowledge Discovery and Data Mining}{August 14--18, 2022}{Washington, DC, USA.}
\acmBooktitle{Proceedings of the 28th ACM SIGKDD Conference on Knowledge Discovery and Data Mining (KDD '22), August 14--18, 2022, Washington, DC, USA}
\acmPrice{15.00}
\acmISBN{978-1-4503-9385-0/22/08}
\acmDOI{10.1145/3534678.3539381}

\begin{document}

\title[Towards Universal Sequence Representation Learning for Recommender Systems]{\texorpdfstring{Towards Universal Sequence Representation Learning\\ for Recommender Systems}{Towards Universal Sequence Representation Learning for Recommender Systems}}

\author{Yupeng Hou$^{*\dagger}$}
\email{houyupeng@ruc.edu.cn}
\affiliation{
    \institution{Gaoling School of Artificial Intelligence, Renmin University of China}
    \country{}
}
\thanks{$*$ Equal contribution.}

\author{Shanlei Mu$^*$}
\email{slmu@ruc.edu.cn}
\affiliation{
    \institution{School of Information, Renmin University of China}
    \country{}
}

\author{Wayne Xin Zhao$^{\dagger\ddagger}$\textsuperscript{\Letter}}
\email{batmanfly@gmail.com}
\affiliation{
    \institution{Gaoling School of Artificial Intelligence, Renmin University of China}
    \country{}
}
\thanks{$\dagger$ Beijing Key Laboratory of Big Data Management and Analysis Methods.}
\thanks{$\ddagger$ Beijing Academy of Artificial Intelligence, Beijing, 100084, China.}
\thanks{\Letter\ Corresponding author.}

\author{Yaliang Li}
\email{yaliang.li@alibaba-inc.com}
\affiliation{
    \institution{Alibaba Group}
    \country{}
}

\author{Bolin Ding}
\email{bolin.ding@alibaba-inc.com}
\affiliation{
    \institution{Alibaba Group}
    \country{}
}

\author{Ji-Rong Wen$^\dagger$}
\email{jrwen@ruc.edu.cn}
\affiliation{
    \institution{Gaoling School of Artificial Intelligence, Renmin University of China}
    \country{}
}

\renewcommand{\authors}{Yupeng Hou, Shanlei Mu, Wayne Xin Zhao, Yaliang Li, Bolin Ding, Ji-Rong Wen}
\renewcommand{\shortauthors}{Hou, et al.}

\newcommand{\ie}{\emph{i.e.,}\xspace}
\newcommand{\eg}{\emph{e.g.,}\xspace}
\newcommand{\aka}{\emph{a.k.a.,}\xspace}
\newcommand{\etal}{\emph{et al.}\xspace}
\newcommand{\paratitle}[1]{\vspace{1.5ex}\noindent\textbf{#1}}
\newcommand{\wrt}{w.r.t.\xspace}
\newcommand{\ignore}[1]{}

\newcommand{\our}{UniSRec}
\newcommand{\ourt}{UniSRec$_{t}$}
\newcommand{\ourid}{UniSRec$_{t+ID}$}

\newcommand{\tba}{\textcolor{red}{xxx }}
\newcommand{\todo}[1]{\textcolor{purple}{#1}}
\newcommand{\tabincell}[2]{\begin{tabular}{@{}#1@{}}#2\end{tabular}}

\begin{abstract}

In order to develop effective sequential recommenders, a series of sequence representation learning (SRL) methods are proposed to model historical user behaviors.
Most existing SRL methods rely on explicit item IDs for developing the sequence models to better capture user preference.
Though effective to some extent, these methods are difficult to be transferred to new recommendation scenarios, due to the limitation by explicitly modeling item IDs.
To tackle this issue, we present a novel universal sequence representation learning approach, named \textbf{UniSRec}.
  The proposed approach utilizes the associated description text of items to learn transferable representations across different recommendation scenarios. For learning \emph{universal item representations}, we design a lightweight item encoding architecture based on parametric whitening and mixture-of-experts enhanced adaptor. For learning \emph{universal sequence representations}, we introduce two contrastive pre-training tasks by sampling multi-domain negatives. With the pre-trained universal sequence representation model, our approach can be effectively transferred to new recommendation domains or platforms in a parameter-efficient way, under either inductive or transductive settings.  Extensive experiments conducted on real-world datasets  demonstrate the effectiveness of the proposed approach. 
  Especially, our approach also leads to a performance improvement in a cross-platform setting, showing the strong transferability of the proposed universal SRL method.
  The code and pre-trained model are available at: \textcolor{blue}{\url{https://github.com/RUCAIBox/UniSRec}}.
\end{abstract}



\maketitle

\section{Introduction}\label{sec:intro}

In the literature of recommender systems, sequential recommendation is a widely studied task~\cite{hidasi2016gru4rec,kang2018sasrec}, aiming to recommend suitable items to a user given her/his historical interaction records. Various methods have been proposed to improve the performance of sequential recommendation, from early matrix factorization (\eg FPMC~\cite{rendle2010fpmc}) to recent sequence neural networks (\eg GRU4Rec~\cite{hidasi2016gru4rec}, Caser~\cite{tang2018caser} and Transformer~\cite{kang2018sasrec,zhou2020s3rec}).
These approaches have largely raised the performance bar of sequential recommendation. 

Though the adopted techniques are different, the core idea of existing methods is similar: it first formulates the user behavior as a chronologically-ordered interaction sequence with the items, and then develops effective architectures for capturing the sequential interaction characteristics that reflect the user preference.
In this way, the learned sequence models can predict the likely items to be interacted by a user, given the observed sequential context. Framed in the paradigm of representation learning~\cite{bengio2013rl}, 
such an approach essentially aims to conduct a sequence representation learning~(SRL) model based on historical user behavior data, which needs to effectively capture the item characteristics and sequential interaction characteristics. 
The capacity of the designed SRL model directly affects the performance of sequential recommendation. 

Despite the progress, most of existing SRL methods for recommendation rely on explicit \emph{item IDs} for developing the sequence models~\cite{hidasi2016gru4rec,kang2018sasrec}. A major issue with this modeling way is that the learned model is difficult to be transferred to new recommendation scenarios, even when the underlying data forms are exactly the same. 
Such an issue limits the reuse of the recommendation model across domains~\cite{zhu2021crossdomain}.
As a common case, we need to re-train a sequential recommender from scratch when adapting to a new domain, which is tedious and resource-consuming. 
In addition, existing sequential recommenders usually suffer from the recommendation of cold-start items having inadequate interactions with users.
For addressing the above issue, a number of studies have been proposed by learning either semantic mapping~\cite{zhu2019dtcdr} or transferable components~\cite{li2022recguru}, in order to bridge the domain gap and enhance the item representations. 
However, these existing attempts cannot fully solve the fundamental issue caused by explicitly modeling item IDs. 
More recently, increasing evidence shows that natural language text can play the role of general semantic bridge 
across different tasks or domains~\cite{devlin2019bert} (including the recommendation tasks, \eg zero-shot recommendation~\cite{ding2021zero}), with the remarkable success of pre-trained language models~(PLM).

Inspired by the recent progress of language intelligence~\cite{vaswani2017attention,devlin2019bert}, we aim to design a new SRL approach for learning more generalizable sequence representations, by breaking the limit of explicit ID modeling. 
The core idea is to utilize the associated description text (\eg product description, product title or brand) of an item, called \emph{item text}, to learn transferable representations across different domains. 
Although previous attempts have shown such an approach is promising~\cite{ding2021zero}, there are still two major challenges to be solved. First, the textual semantic space is not directly suited for the recommendation tasks. It is not clear how to model and utilize item texts for improving the recommendation performance, since directly introducing raw textual representations as additional features may lead to suboptimal results. Second, it is difficult to leverage multi-domain data for improving the target domain, where the \emph{seesaw phenomenon} (referring to learning from multiple kinds of domain-specific patterns is conflict or oscillating) often appears~\cite{tang2020ple}.

To address the above issues, we propose the universal sequence representation learning approach, named \textbf{\our}. 
Our approach takes general interaction sequences as input, and learns universal ID-agnostic representations based on a pre-training approach. Specially, we focus on two key points that learn \emph{universal item representation} and \emph{universal sequence representations}.
 For learning universal item representations, we design a lightweight architecture based on parametric whitening and mixture-of-experts enhanced adaptor, which can derive more isotropic semantic representations as well as enhance the domain fusion and adaptation. 
 For learning universal sequence representations, we introduce two kinds of contrastive learning tasks, namely sequence-item and sequence-sequence contrastive tasks, by sampling multi-domain negatives. Based on the above methods, the pre-trained model can be effectively transferred to a new recommendation scenario in a parameter-efficient way, under either inductive or transductive settings.

To evaluate the proposed approach UniSRec, we conduct extensive experiments on real-world datasets from different application domains and platforms.
Experimental results demonstrate that
the proposed approach can effectively utilize data from multiple domains to learn universal and transferable representations. 
Especially, the results on cross-platform experiments show that recommendation performance can be improved with the universal sequence representation model pre-trained on other platforms without overlapping users or items.

\section{Methodology}

In this section, we present the proposed \textbf{Uni}versal \textbf{S}equence representation learning approach for \textbf{Rec}ommendation, named as \textbf{\our}.
Given historical user behavioral sequences from a mixture of multiple domains, we aim to learn universal item and sequence representations that can effectively transfer and generalize to new recommendation scenarios (\eg new domains or platforms) in a parameter-efficient way.

\subsection{Overview of the Approach}

Our approach takes general interaction sequences as input and learns universal representations based on a pre-training approach. 
We then formulate the task and overview the proposed approach.

\paratitle{General input formulation}. We  formulate the behavior sequence of a user in a general form of interaction sequence $s = \{ i_1, i_2, \cdots i_n \}$, 
 (in a chronological order of interaction time), where each interacted item $i$  
is associated with a unique item ID and a description text (\eg the product description, item title or brand). We call the description text of an item $i$ \emph{item text}, denoted by $t_{i}=\{w_1, w_2, \cdots, w_c \}$, where the words $w_j$ are from a shared vocabulary and $c$ denotes the truncated length of item text.
Here, each  sequence contains all the interaction behavior of a user at some specific domain, and a user can generate multiple behavior sequences at different domains or platforms.
As discussed before, as there are large semantic gap between different domains, we don't simply mix the behavior data of a user. Instead, we treat the multiple interaction sequences of a user as different sequences, without explicitly maintaining user IDs for each sequence.   
Note that unlike other pre-training based recommendation methods~\cite{zhou2020s3rec,yuan2020peterrec}, item IDs are only auxiliary information in our approach, and we mainly utilize item text to derive generalizable ID-agnostic representations. Unless specified, item IDs will not be used as input of our approach.

\begin{figure*}[ht]
    \centering
    \includegraphics[width=0.82\textwidth]{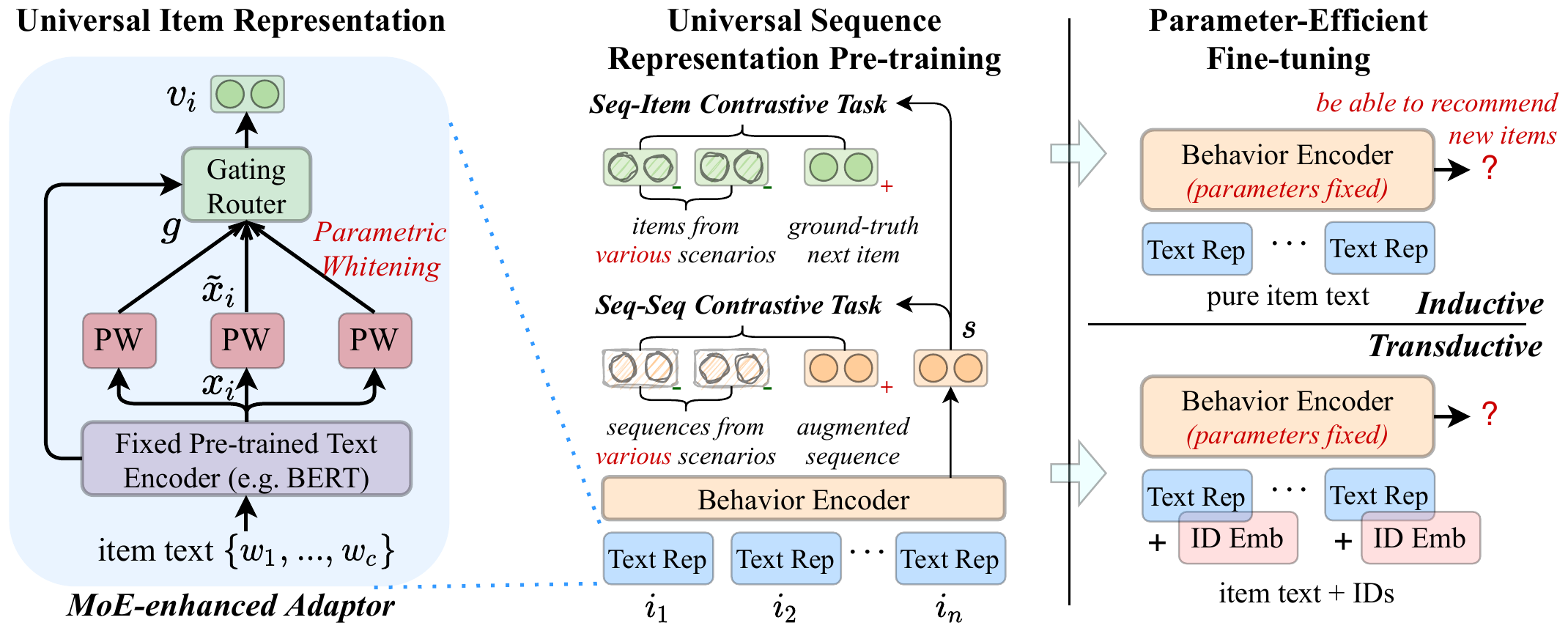}
  \caption{The overall framework of the proposed universal sequence representation learning approach (UniSRec).}
    \label{fig:overall}
\end{figure*}

\paratitle{Solutions}. To learn transferable representations across domains, 
we identify two key problems for achieving this purpose, \ie learning universal item representation and sequence representation, since \emph{items} and \emph{sequences} are the basic data forms in our general formulation. For learning universal item representations (Section 2.2), we focus on the domain fusion and adaptation with an MoE-enhanced adaptor based on parametric whitening. For learning universal sequence representations (Section 2.3), we introduce two kinds of contrastive learning tasks, namely sequence-item and sequence-sequence contrastive tasks, by sampling multi-domain negatives. Based on the above methods, the pre-trained model can be effectively transferred to a new recommendation scenario in a parameter-efficient way, under either inductive or transductive settings (Section 2.4). 
The overall framework of the proposed approach \our~is depicted in Figure~\ref{fig:overall}.

\subsection{Universal Textual Item Representation}\label{sec:uni_item_rep}

The first step toward universal sequential behavior modeling is to represent items from various recommendation scenarios (\eg domains or platforms) into a unified semantic space. 
In previous studies~\cite{hidasi2016gru4rec,kang2018sasrec}, item representations are usually learned in 
\emph{transductive learning} setting, where item IDs are pre-given and ID embeddings are learned as item representations. Such a way largely limits the transferability of item representations, since the vocabularies of item IDs are usually different across domains.

Our solution is to learn transferable item representations based on the associated item text, which describes the item characteristics in the form of natural language. 
As more and more evidence shows~\cite{devlin2019bert}, natural language provides a general data form to bridge the semantic gap across different tasks or domains. 
Based on this idea, we first utilize the pre-trained language model~(PLMs) to learn the text embeddings. Since the derived text representations from different domains are likely to span different semantic spaces (even with the same text encoder), we propose the techniques of 
parametric whitening and mixture-of-experts (MoE) enhanced adaptor to transform the text semantics into a universal form suited to the recommendation tasks.

\subsubsection{Textual Item Encoding via Pre-trained Language Model}
Considering the excellent language modeling capacity of PLMs, we utilize the widely used BERT model~\cite{devlin2019bert} to learn universal text representations for representing items. 
 Given an item $i$ and its corresponding text $t_i$, we firstly concatenate (1) a special symbol \texttt{[CLS]}, (2) the words of item text $\{w_1, w_2, \ldots, w_c\}$, in order and derive the input sequence for BERT. Then we feed the concatenated sequence into the BERT model, and we have:
\begin{equation}
    \bm{x}_i = \operatorname{BERT}([\texttt{[CLS]}; w_1, \ldots, w_c]),
\end{equation}
where $\bm{x}_i \in \mathbb{R}^{d_W}$ is the final hidden vector corresponding to the first input token (\texttt{[CLS]}), and ``[;]'' denotes the concatenation operation.

\subsubsection{Semantic Transformation via Parametric Whitening}

Though we can obtain semantic representations from BERT, they are not directly suited for the recommendation tasks.
Existing studies~\cite{li2020bertflow} have found that BERT induces a non-smooth anisotropic semantic space for general texts. The case will become more severe when we mix item texts from multiple domains with a large semantic gap. 
Inspired by recent works on whitening-based methods~\cite{su2021whitening,huang2021white2}, we conduct a simple linear transformation to transform original BERT representations for deriving isotropic semantic representations.

Different from the original whitening methods~\cite{su2021whitening,huang2021white2} with pre-set mean and variance,
we incorporate learnable parameters in the whitening transformation for better generalizability on unseen domains.
 Formally, we have:
\begin{equation}
   \widetilde{\bm{x}}_i = (\bm{x}_i-\bm{b}) \cdot \bm{W}_1,\label{eq:whitening}
\end{equation}
where $\bm{b} \in \mathbb{R}^{d_W}$ and $\bm{W}_1 \in \mathbb{R}^{d_W \times d_V}$ are learnable parameters, and $\widetilde{\bm{x}}_i \in \mathbb{R}^{d_V}$ is the transformed representation.
In this way, the anisotropy issue of learned representations can be alleviated, which is useful to learn universal semantic representations.
For the efficiency consideration, we don't introduce complex non-linear architectures, such as flow-based generative models~\cite{li2020bertflow}, which will be studied in future work.

\subsubsection{Domain Fusion and Adaptation via MoE-enhanced Adaptor}\label{sec:adaptor}

With the above whitening transformation, our model can learn more isotropic semantic representations. 
In order to learn universal item representations, another important issue is how to transfer and fuse information across domains, since there is usually a large semantic gap between different domains. For example,the top frequent words of item text are quite different across domains, \eg \emph{natural}, \emph{sweet}, \emph{fresh} for food domain and \emph{war}, \emph{love}, \emph{story} for movies domain. %
A straightforward approach is to map the original BERT embeddings into some shared semantic space~\cite{li2020bertflow}. However, it  will lead to limited representation capacity for migrating  the domain bias.   
As our solution, we learn multiple whitening embeddings for an item, and utilize an adaptive combination of these embeddings as the universal item representations. Not limited to a simple single mapping between item representations, 
we aim to establish a more flexible representation mechanism to capture the semantic relatedness  for domain fusion and adaptation.

To implement our idea, we employ the mixture-of-expert (MoE) architecture~\cite{shazeer2017moe} for learning more generalizable item representations. Specially, we 
 incorporate $G$ whitening transformation modules  as the \emph{experts}, and then construct the MoE-enhanced adaptor based on a parameterized router:
\begin{equation}
    \bm{v}_i = \sum_{k=1}^{G} g_k \cdot \widetilde{\bm{x}}_i^{(k)},\label{eq:h_i}
\end{equation}
where $\widetilde{\bm{x}}_i^{(k)}$ is the output of the $k$-th whitening transformation module (Eqn.~\eqref{eq:whitening}) and  $g_k$ is the corresponding combination weight from the \emph{gating router}, defined as follows:
\begin{align}
    \bm{g} =& \operatorname{Softmax}\left(\bm{x}_i \cdot \bm{W}_2 +   \bm{\delta} \right),\\ \label{eq:router}
    \bm{\delta} =& \operatorname{Norm}() \cdot \operatorname{Softplus}\left(\bm{x}_i \cdot \bm{W}_{3}\right).
\end{align}
In this equation, we utilize original BERT embedding $\bm{x}_i$ as input of the router module, 
as it contains domain-specific semantic bias. Furthermore, we incorporate learnable parameter matrices $\bm{W}_2,\bm{W}_3\in \mathbb{R}^{d_W \times G}$  to adaptively adjust  the weights of experts $\bm{g} \in \mathbb{R}^{G}$. In order to balance the expert load, we utilize $\operatorname{Norm}()$ to generate random Gaussian noise,  controlled by parameter $\bm{W}_2$. %

The merits of the MoE-enhanced adaptor are threefold. Firstly,  the representation of a single item is enhanced by learning multiple whitening transformations.  
Second, 
we no longer require a direct semantic mapping across domains, but instead utilize a learnable gating mechanism to adaptively establish the semantic relatedness for domain fusion and adaptation. 
Third, the lightweight adaptor endows the flexibility of parameter-efficient fine-tuning when adapting to new domains (detailed in Section~\ref{sec:finetune}).

\subsection{Universal Sequence Representation}\label{sec:uni_seq_rep}

Since different domains usually correspond to varying user behavioral patterns, it may not work well to simply mix interaction sequences from multiple domains for pre-training. And, it is likely to lead to the \emph{seesaw phenomenon}~\cite{tang2020ple} that the learning from multiple domain-specific behavioral patterns can be conflict. 
Our solution is to introduce two kinds of contrastive learning tasks, which can further enhance the fusion and adaptation of different domains in deriving the item representations. 
In what follows, we firstly introduce the base behavior encoder architecture, and then present the proposed contrastive pre-training tasks that enhance the sequence representations in universal semantic space.

\subsubsection{Self-attentive Sequence Encoding}

Given a sequence of universal item representations, we further utilize a user behavior encoder to obtain the sequence representation. We aim to construct the sequential patterns based on the learned universal textual item representations, but not item IDs.
Here, we adopt a widely used self-attentive architecture~\cite{vaswani2017attention}, \ie Transformers.
Specially, it consists stacks of multi-head self-attention layers (denoted by $\operatorname{MHAttn}(\cdot)$) and point-wise feed-forward networks (multilayer perceptron activated by $\operatorname{ReLU}$, denoted by $\operatorname{FFN}(\cdot)$). We sum the learned text representations (\ie $\bm{v}_i$ in Eqn.~\eqref{eq:h_i}) and the absolute position embeddings $\bm{p}_j$ as input at position $j$. The input and the update process can be formalized as following:
\begin{align}
    \bm{f}^{0}_j =& \ \bm{v}_i + \bm{p}_j,\\ \label{eq:input}
    \bm{F}^{l+1} =& \operatorname{FFN}(\operatorname{MHAttn}(\bm{F}^l)),
\end{align}
where $\bm{F}^l = [\bm{f}^l_0;\ldots;\bm{f}^l_n]$ denotes the concatenated representations at each position in the $l$-th layer.
We take the final hidden vector $\bm{f}_{n}^{L}$ corresponding to the $n$-th (last) position as the sequence representation (a total number of  $L$ layers in the behavior encoder).

\subsubsection{Multi-domain Sequential Representation Pre-training}

Given interaction sequences from multiple domains, we next study 
 how to design suitable optimization objectives to derive the outputs of sequential encoder in the unified representation space. By contrasting sequences and items from different domains, we aim to alleviate the seesaw phenomenon and capture their semantic correlation in the pre-training stage. For this purpose, we design the following sequence-item and sequence-sequence contrastive tasks.

\paratitle{Sequence-item contrastive task.} The sequence-item contrastive task aims to capture the intrinsic correlation 
between sequential contexts (\ie the observed subsequence) and potential next items in an interaction sequence. 
Different from previous next-item prediction task~\cite{kang2018sasrec,lin2022ncl} (using \emph{in-domain} negatives), for a given sequence, we adopt \emph{across-domain} items as negatives. Such a way can enhance both the semantic fusion and adaptation across domains, which is helpful to learn universal sequence representations. 

We consider a batch setting of $B$ training instances, where each training instance is a pair of the sequential context (containing the proceeding items) and the positive next item. We first encode them   
 into embedding representations $\{ \langle \bm{s}_{1}, \bm{v}_{1}\rangle, \ldots, \langle\bm{s}_{B}, \bm{v}_{B}\rangle\}$, where $\bm{s}$ represents the normalized contextual sequence representations, and $\bm{v}$ denotes the representation of the  positive next item.
Then, we formalize the sequence-item contrastive loss as follows:
\begin{equation}
    \ell_{S-I} = -\sum_{j=1}^{B} \log \frac{\exp{\left(\bm{s}_j\cdot\bm{v}_j/\tau\right)}}{\sum_{j'=1}^{B} \exp{\left(\bm{s}_j\cdot\bm{v}_{j'}/\tau\right)}},\label{eq:loss_si}
\end{equation}
where in-batch items are regarded as negative instances and $\tau$ is a temperature parameter. As batches are constructed randomly, the in-batch negative instances $\{\bm{v}_{j'}\}$ will contain items from a mixture of multiple domains. 

\paratitle{Sequence-sequence contrastive task.} 
Besides the above item-level pre-training task, we further propose a sequence-level pre-training task, by conducting the 
contrastive learning among multi-domain interaction sequences. 
The object is to discriminate the representations of augmented sequences from multi-domain sequences.
We consider two kinds of augmentation strategies: (1) \emph{Item drop} refers to randomly dropping a fixed ratio of items in the original sequence, and (2) \emph{Word drop} refers to randomly dropping words in item text. Given a target sequence (with the representation $\bm{s}_j$), the augmented ones are considered as positives (with the representation $\widetilde{\bm{s}}_{j}$), while other in-batch ones are considered as negatives.   
The sequence-sequence contrastive loss can be formally presented as following:
\begin{equation}
    \ell_{S-S} = -\sum_{j=1}^{B} \log \frac{\exp{\left(\bm{s}_j\cdot \widetilde{\bm{s}}_{j}/\tau\right)}}{\sum_{j'=1}^{B} \exp{\left(\bm{s}_j\cdot\bm{s}_{j'}/\tau\right)}}.\label{eq:loss_ss}
\end{equation}

Similar to Eqn.~\eqref{eq:loss_si}, as batches are constructed randomly, the in-batch negative instances  naturally contain sequences from multiple domains. 
In the implementation, we preprocess the augmented item text using \emph{word drop} for efficient pre-training, as the BERT representations of item text can be obtained during preprocessing.

\paratitle{Multi-task learning.} At the pre-training stage, we leverage a multi-task training strategy to jointly optimize the proposed sequence-item contrastive loss in Eqn.~\eqref{eq:loss_si} and sequence-sequence contrastive loss in Eqn.~\eqref{eq:loss_ss}:
\begin{equation}
    \mathcal{L}_{\text{PT}} = \ell_{S-I} + \lambda \cdot \ell_{S-S},
\end{equation}
where $\lambda$ is a hyper-parameter to control the weight of sequence-sequence contrastive loss.
The pre-trained model is to be fine-tuned for adapting to new domains.

\subsection{Parameter-Efficient Fine-tuning}\label{sec:finetune}

To adapt to a new domain, previous pre-training based recommendation methods~\cite{zhou2020s3rec,ding2021zero} 
usually require fine-tuning the whole network architecture, which is  time-consuming and less flexible. Since our model can learn universal representations for interaction sequences, our idea is to fix the parameters of the major architecture, while only fine-tuning a small proportion of parameters from the MoE-enhanced adaptor (Section~\ref{sec:adaptor}) for incorporating necessary adaptation. 
 We find that the proposed MoE-enhanced adaptor can quickly adapt to  unseen domains, fusing the pre-trained model with new domain characteristics. 
To be specific, we consider two fine-tuning settings, either \emph{inductive} or \emph{transductive}, based on whether  item IDs in the target domain can be accessed.

\paratitle{Inductive setting.} The first setting considers the test cases of recommending new items from an unseen domain, which can't be well solved by an ID-based recommendation model.  
The proposed model doesn't rely on item IDs, so that it can learn universal text representations for new items. Given a training sequence from the target domain, we firstly encode the sequential context ($i_1 \rightarrow i_t$) and the candidate item $i_{t+1}$ into 
universal representations as $\bm{s}$ and $\bm{v}_{i_{t+1}}$. 
Then we predict the next item according to the following probability:
\begin{equation}
    P_I( i_{t+1} | s) = \operatorname{Softmax}(\bm{s} \cdot \bm{v}_{i_{t+1}}),\label{eq:prob_i}
\end{equation}
where we compute the softmax probability over the candidate set (the positive item and a number of sampled negatives).
The parameters to be tuned are those in $\bm{b}$ and $\bm{W}_r$ in Eqn.~\eqref{eq:whitening}. 

\paratitle{Transductive setting.} The second setting assumes that nearly all the items of the target domain have appeared in the training set, and we can also learn ID embeddings since item IDs are available.
In this setting, to represent an item, we combine the textual embedding $\bm{v}_{i_{t+1}}$ and ID embedding $\bm{e}_{i_{t+1}}$ as the final item representation. 
 Thus, we have the following prediction probability:
\begin{equation}
    P_T(i_{t+1} | s) = \operatorname{Softmax}\left(\widetilde{\bm{s}} \cdot (\bm{v}_{i_{t+1}} + \bm{e}_{i_{t+1}} )\right),\label{eq:prob_t}
\end{equation}
where $\widetilde{\bm{s}}$ denotes the enhanced universal sequence representation by adding ID embeddings in the input (Eqn.~\eqref{eq:input}). Note that the rest parameters of the sequence encoder are still fixed in this setting.

For each setting, we optimize the widely used cross-entropy loss to fine-tune parameters of MoE-enhanced adaptors. After fine-tuning, we predict the probability distribution of next item for given sequences via Eqn.~\eqref{eq:prob_i} and Eqn.~\eqref{eq:prob_t}.

\begin{table}[t]
\small
\caption{Comparison of the transfer learning scenarios and application settings of several approaches. $1\rightarrow 1$ denotes $1$ source domain to $1$ target domain, and $M \rightarrow N$ denotes $M$ source domains to $N$ target domains. ``Non-OL'' denotes that the approach doesn't require overlapped users or items.}
\label{tab:cmp}
\resizebox{\columnwidth}{!}{
\begin{tabular}{@{}lccccc@{}}
\toprule
\multirow{2}{*}{Methods} & \multicolumn{3}{c}{Transfer Learning Scenarios} & \multicolumn{2}{c}{Application Settings} \\ \cmidrule(l){2-4} \cmidrule(l){5-6}
          & $1 \rightarrow 1$ & $M \rightarrow N$ & Non-OL & Transductive            & Inductive            \\ \midrule
S$^3$-Rec~\cite{zhou2020s3rec}    & \textcolor{purple}{\XSolidBrush} & \textcolor{purple}{\XSolidBrush} & \textcolor{purple}{\XSolidBrush} & \textcolor{teal}{\CheckmarkBold} & \textcolor{purple}{\XSolidBrush} \\ 
PeterRec~\cite{yuan2020peterrec}  & \textcolor{teal}{\CheckmarkBold} & \textcolor{purple}{\XSolidBrush} & \textcolor{purple}{\XSolidBrush} & \textcolor{teal}{\CheckmarkBold} & \textcolor{purple}{\XSolidBrush} \\ 
RecGURU~\cite{li2022recguru}   & \textcolor{teal}{\CheckmarkBold} & \textcolor{purple}{\XSolidBrush} & \textcolor{teal}{\CheckmarkBold} & \textcolor{teal}{\CheckmarkBold} & \textcolor{purple}{\XSolidBrush} \\ 
ZESRec~\cite{ding2021zero} & \textcolor{teal}{\CheckmarkBold} & \textcolor{purple}{\XSolidBrush} & \textcolor{teal}{\CheckmarkBold} & \textcolor{purple}{\XSolidBrush} & \textcolor{teal}{\CheckmarkBold} \\ 
UniSRec (ours) & \textcolor{teal}{\CheckmarkBold} & \textcolor{teal}{\CheckmarkBold} & \textcolor{teal}{\CheckmarkBold} & \textcolor{teal}{\CheckmarkBold} & \textcolor{teal}{\CheckmarkBold} \\ \bottomrule
\end{tabular}
}
\end{table}

\subsection{Discussion}\label{sec:discuss}

In the literature of recommender systems, a large number of recommendation models have been developed. Here, we make a brief comparison with  related recommendation models, in order to highlight the novelty and differences of our approach.

\textbf{General sequential approaches} such as GRU4Rec~\cite{hidasi2016gru4rec} and SASRec~\cite{kang2018sasrec} 
rely on explicit item IDs to construct the sequential model, where they assume item IDs are pre-given. These approaches can't perform well under the cold-start setting with new items. As a comparison, our approach aims to construct an ID-agnostic recommendation model that can capture sequential patterns in a more general form of natural language.

\textbf{Cross-domain approaches} such as RecGURU~\cite{li2022recguru} propose to leverage the auxiliary information from source domains to improve the performance on a target domain. However, most approaches require overlapping users or items as anchors. Besides, it is not easy to transfer and fuse multiple source domains for improving the target domain. For our approach, we propose an MoE-enhanced adaptor mechanism for domain fusion and adaptation based on universal textual semantics. 

\textbf{Pre-training sequential approaches} mainly pre-train their model on sequences from 
 (1) the current domain (S$^3$-Rec~\cite{zhou2020s3rec}, IDA-SR~\cite{mu2022ida}),
 (2) other domains with overlapping users (PeterRec~\cite{yuan2020peterrec})
 or (3) other closely related domains (ZESRec~\cite{ding2021zero}). However, none of these methods explore how to pre-train universal item and sequence representations on multiple weakly related or irrelevant domains. 
 In contrast, our approach can learn more transferable representations  that well generalize to the target domains, from a mixture of multiple source domains.  
 The comparison of these approaches is presented in Table~\ref{tab:cmp}.

\section{Experiments}

In this section, we first set up the experiments, and then present  the results and analysis.

\subsection{Experimental Setup}

\begin{table}[!t] %
	\caption{Statistics of the datasets after preprocessing. ``Avg. $n$'' denotes the average length of item sequences. ``Avg. $c$'' denotes the average number of tokens in item text.
	}
	\label{tab:dataset}
	\resizebox{\columnwidth}{!}{
	\begin{tabular}{l *{5}{r}}
		\toprule
		\textbf{Datasets} & \textbf{\#Users} & \textbf{\#Items} & \textbf{\#Inters.} & \textbf{Avg. $n$} & \textbf{Avg. $c$}\\
		\midrule
		\textbf{Pre-trained} & 1,361,408 & 446,975 & 14,029,229 & 13.51 & 139.34 \\
		- Food   & 115,349 &  39,670 & 1,027,413 &  8.91 & 153.40 \\
		- CDs    &  94,010 &  64,439 & 1,118,563 & 12.64 & 80.43 \\
		- Kindle & 138,436 &  98,111 & 2,204,596 & 15.93 & 141.70 \\
		- Movies & 281,700 &  59.203 & 3,226,731 & 11.45 & 97.54 \\
		- Home   & 731,913 & 185,552 & 6,451,926 &  8.82 & 168.89 \\
		\midrule
		\textbf{Scientific}  &  8,442 &  4,385 &  59,427 & 7.04 & 182.87 \\
		\textbf{Pantry}      & 13,101 &  4,898 & 126,962 & 9.69 & 83.17 \\
		\textbf{Instruments} & 24,962 &  9,964 & 208,926 & 8.37 & 165.18 \\
		\textbf{Arts}        & 45,486 & 21,019 & 395,150 & 8.69 & 155.57 \\
		\textbf{Office}      & 87,436 & 25,986 & 684,837 & 7.84 & 193.22 \\ \midrule
		\textbf{Online Retail} & 16,520 & 3,469 & 519,906 & 26.90 & 27.80 \\
		\bottomrule
	\end{tabular}
	}
\end{table}

\begin{table*}[!ht]
\centering
\caption{Performance comparison of different recommendation models. The best and the second-best performances are denoted in bold and underlined fonts, respectively. ``Improv.'' indicates the relative improvement ratios of the proposed approach over the best performance baselines. ``*'' denotes that the improvements are significant at the level of 0.01 with paired $t$-test.}
\label{tab:exp-main}
\resizebox{2.1\columnwidth}{!}{
\begin{tabular}{@{}ccccccccccccr@{}}
\toprule
Scenario & Dataset & Metric & SASRec  & BERT4Rec & FDSA & S$^3$-Rec & CCDR & RecGURU & ZESRec & \ourt & \ourid & Improv. \\ \midrule \midrule
\multirow{20}{*}{\shortstack{Cross-\\Domain}} &
\multirow{4}{*}{Scientific} &
     Recall@10 & 0.1080 & 0.0488 & 0.0899 & 0.0525 & 0.0695 & 0.1023 & 0.0851 & \underline{0.1188}* & \textbf{0.1235}* & +14.35\% \\
 & & NDCG@10   & 0.0553 & 0.0243 & 0.0580 & 0.0275 & 0.0340 & 0.0572 & 0.0475 & \textbf{0.0641}* & \underline{0.0634}* & +10.52\% \\
 & & Recall@50 & 0.2042 & 0.1185 & 0.1732 & 0.1418 & 0.1647 & 0.2022 & 0.1746 & \underline{0.2394}* & \textbf{0.2473}* & +21.11\% \\
 & & NDCG@50   & 0.0760 & 0.0393 & 0.0759 & 0.0468 & 0.0546 & 0.0786 & 0.0670 & \underline{0.0903}* & \textbf{0.0904}* & +15.01\% \\
\cmidrule(l){2-13}
 & \multirow{4}{*}{Pantry} &
     Recall@10 & 0.0501 & 0.0308 & 0.0395 & 0.0444 & 0.0408 & 0.0469 & 0.0454 & \underline{0.0636}* & \textbf{0.0693}* & +38.32\% \\
 & & NDCG@10   & 0.0218 & 0.0152 & 0.0209 & 0.0214 & 0.0203 & 0.0209 & 0.0230 & \underline{0.0306}* & \textbf{0.0311}* & +35.22\% \\
 & & Recall@50 & 0.1322 & 0.1030 & 0.1151 & 0.1315 & 0.1262 & 0.1269 & 0.1141 & \underline{0.1658}* & \textbf{0.1827}* & +38.20\% \\
 & & NDCG@50   & 0.0394 & 0.0305 & 0.0370 & 0.0400 & 0.0385 & 0.0379 & 0.0378 & \underline{0.0527}* & \textbf{0.0556}* & +39.00\% \\
 \cmidrule(l){2-13}
 & \multirow{4}{*}{Instruments} &
     Recall@10 & 0.1118 & 0.0813 & 0.1070 & 0.1056 & 0.0848 & 0.1113 & 0.0783 & \underline{0.1189}* & \textbf{0.1267}* & +13.33\% \\
 & & NDCG@10   & 0.0612 & 0.0620 & \textbf{0.0796} & 0.0713 & 0.0451 & 0.0681 & 0.0497 & 0.0680 & \underline{0.0748}* & $-$ \\
 & & Recall@50 & 0.2106 & 0.1454 & 0.1890 & 0.1927 & 0.1753 & 0.2068 & 0.1387 & \underline{0.2255}* & \textbf{0.2387}* & +13.34\% \\
 & & NDCG@50   & 0.0826 & 0.0756 & \underline{0.0972} & 0.0901 & 0.0647 & 0.0887 & 0.0627 & 0.0912 & \textbf{0.0991}* & +1.95\% \\
 \cmidrule(l){2-13}
 & \multirow{4}{*}{Arts} &
     Recall@10 & \underline{0.1108} & 0.0722 & 0.1002 & 0.1003 & 0.0671 & 0.1084 & 0.0664 & 0.1066 & \textbf{0.1239}* & +11.82\% \\
 & & NDCG@10   & 0.0587 & 0.0479 & \textbf{0.0714} & 0.0601 & 0.0348 & 0.0651 & 0.0375 & 0.0586 & \underline{0.0712} & $-$ \\
 & & Recall@50 & 0.2030 & 0.1367 & 0.1779 & 0.1888 & 0.1478 & 0.1979 & 0.1323 & \underline{0.2049}* & \textbf{0.2347}* & +15.62\% \\
 & & NDCG@50   & 0.0788 & 0.0619 & \underline{0.0883} & 0.0793 & 0.0523 & 0.0845 & 0.0518 & 0.0799 & \textbf{0.0955}* & +8.15\% \\
 \cmidrule(l){2-13}
 & \multirow{4}{*}{Office} &
     Recall@10 & 0.1056 & 0.0825 & 0.1118 & 0.1030 & 0.0549 & \underline{0.1145} & 0.0641 & 0.1013 & \textbf{0.1280}* & +11.79\% \\
 & & NDCG@10   & 0.0710 & 0.0634 & \textbf{0.0868} & 0.0653 & 0.0290 & 0.0768 & 0.0391 & 0.0619 & \underline{0.0831} & $-$ \\
 & & Recall@50 & 0.1627 & 0.1227 & 0.1665 & 0.1613 & 0.1095 & \underline{0.1757} & 0.1113 & 0.1702 & \textbf{0.2016}* & +14.74\% \\
 & & NDCG@50   & 0.0835 & 0.0721 & \underline{0.0987} & 0.0780 & 0.0409 & 0.0901 & 0.0493 & 0.0769 & \textbf{0.0991} & +0.41\% \\
 \midrule
\multirow{4}{*}{\shortstack{Cross-\\Platform}} &
\multirow{4}{*}{\shortstack{Online\\ Retail}} &
     Recall@10 & 0.1460 & 0.1349 & \underline{0.1490} & 0.1418 & 0.1347 & 0.1467 & 0.1103 & 0.1449 & \textbf{0.1537}* & +3.15\% \\
 & & NDCG@10   & 0.0675 & 0.0653 & \underline{0.0719} & 0.0654 & 0.0620 & 0.0658 & 0.0535 & 0.0677 & \textbf{0.0724} & +0.70\% \\
 & & Recall@50 & 0.3872 & 0.3540 & 0.3802 & 0.3702 & 0.3587 & \textbf{0.3885} & 0.2750 & 0.3604 & \textbf{0.3885} & 0.00\% \\
 & & NDCG@50   & 0.1201 & 0.1131 & \underline{0.1223} & 0.1154 & 0.1108 & 0.1188 & 0.0896 & 0.1149 & \textbf{0.1239}* & +1.31\% \\
 \bottomrule
\end{tabular}
}
\end{table*}

\subsubsection{Datasets}\label{sec:dataset}

To evaluate the performance of the proposed approach, we conduct experiments in both cross-domain setting and cross-platform setting.
The statistics of datasets after preprocessing are summarized in Table~\ref{tab:dataset}.

(1) \textbf{Pre-trained datasets}: we select five categories from Amazon review datasets~\cite{ni2019amazon}, ``\emph{Grocery and Gourmet Food}'', ``\emph{Home and Kitchen}'', ``\emph{CDs and Vinyl}'', ``\emph{Kindle Store}'' and ``\emph{Movies and TV}'', as the source domain datasets for pre-training.

(2) \textbf{Cross-domain datasets}: we select another five categories from Amazon review datasets~\cite{ni2019amazon}, ``\emph{Prime Pantry}'', ``\emph{Industrial and Scientific}'', ``\emph{Musical Instruments}'', ``\emph{Arts, Crafts and Sewing}'' and ``\emph{Office Products}'', as target domain datasets to evaluate the proposed approach in cross-domain setting. 

(3) \textbf{Cross-platform datasets}: we also select a dataset from different platforms to evaluate the pre-trained universal sequence representation model in a cross-platform setting. \textbf{Online Retail}\footnote{https://www.kaggle.com/carrie1/ecommerce-data} contains transactions occurring between 01/12/2010 and 09/12/2011 from a UK-based online retail platform, which does not contain shared users or items with the Amazon platform. 

Following previous works~\cite{kang2018sasrec,zhou2020s3rec}, we keep the five-core datasets and filter users and items with fewer than five interactions for all datasets.
Then we group the interactions by users and sort them by timestamp ascendingly.
For item text, we concatenate fields including \emph{title}, \emph{categories} and \emph{brand} in Amazon dataset and 
directly use the \emph{Description} field in Online Retail dataset.
We truncate item text longer than $512$ tokens.

\subsubsection{Compared Methods}
We compare the proposed approach with the following baseline methods:

$\bullet$ \textbf{SASRec}~\cite{kang2018sasrec} adopts a self-attention network to capture the user's preference within a sequence.

$\bullet$ \textbf{BERT4Rec}~\cite{sun2019bert4rec} adapts the original text-based BERT model with the cloze objective 
for modeling user behavior sequences. 

$\bullet$ \textbf{FDSA}~\cite{zhang2019fdsa} proposes to capture item and feature transition patterns via self-attentive networks.

$\bullet$ \textbf{S$^3$-Rec}~\cite{zhou2020s3rec} pre-trains sequential models via mutual information maximization objectives for feature fusion. 

$\bullet$ \textbf{CCDR}~\cite{xie2021ccdr} proposes intra-domain and inter-domain contrastive objects for cross-domain recommendation in matching. We extract textual tags using TF-IDF algorithm, and then optimize the taxonomy-based inter-CL objective. 

$\bullet$ \textbf{RecGURU}~\cite{li2022recguru} proposes to pre-train user representations via autoencoder in an adversarial learning paradigm. In our implementation, we remove constraints on overlapped users. 

$\bullet$ \textbf{ZESRec}~\cite{ding2021zero} encodes item text via pre-trained language model as item representations. ZESRec can be pre-trained on source domain and directly applied to target domains for zero-shot recommendation. For a fair comparison, we fine-tune the pre-trained model using item sequences of the target domain.

For our approach, we firstly pre-train a universal sequence representation model on the five source datasets that are introduced in Section~\ref{sec:dataset}.
We consider two major variants: (1) \textbf{\underline{\ourt}} denotes the model fine-tuned in inductive setting using only item text; (2)  \textbf{\underline{\ourid}} denotes the model fine-tuned in transductive setting using both item ID and item text.

\subsubsection{Evaluation Settings}
To evaluate the performance of the next item prediction task, we adopt two widely used metrics Recall@$N$ and NDCG@$N$, where $N$ is set to 10 and 50. Following previous works~\cite{sun2019bert4rec,zhou2020s3rec,zhao2020revisiting}, we apply the leave-one-out strategy for evaluation. For each user interaction sequence, the last item is used as the test data, the item before the last one is used as the validation data, and the remaining interaction records are used for training. 
We rank the ground-truth item of each sequence among all the other items for evaluation on test set, and finally report the average score of all test users.

\subsubsection{Implementation Details}

We implement UniSRec using a popular open-source recommendation library \textsc{RecBole}\footnote{\url{https://recbole.io}}~\cite{zhao2021recbole}.
To ensure a fair comparison, we optimize all the methods with Adam optimizer and carefully search the hyper-parameters of all the compared methods. The batch size is set to 2,048. We adopt early stopping with the patience of 10 epochs to prevent overfitting, and NDCG@10 is set as the indicator. We tune the learning rate in $\{0.0003, 0.001, 0.003, 0.01\}$ and the embedding dimension in $\{64, 128, 300\}$.
We pre-train the proposed approach for $300$ epochs with $\lambda=1e^{-3}$ and $G=8$ experts. 

\subsection{Overall Performance}

We compare the proposed approach  with the baseline methods on the five cross-domain datasets and one cross-platform dataset. Note that, we fine-tune the same pre-trained universal sequence representation model on these six datasets for our approach. The results are reported in Table~\ref{tab:exp-main}.

For the baseline methods, text-enhanced sequential recommendation methods (\ie FDSA and S$^3$-Rec) perform better than the traditional sequential recommendation methods (\ie SASRec and BERT4Rec) on several datasets, since item texts are used as auxiliary features to improve the performance. The cross-domain methods CCDR and RecGURU do not perform well, because they are still in a transductive setting. These methods 
become less effective without explicit overlapping users between source and target domains. ZESRec utilizes PLMs to encode item texts, but mainly reuses existing architectures and pre-training tasks, which  can't fully leverage multi-domain interaction data for improving the target domain.

Finally, by comparing the proposed approach \ourid~ with all the baselines, it is clear that \ourid~achieves the best performance in almost all the cases. Different from these baselines, we derive universal sequence representations via pre-training on multi-domain datasets. With the specially designed parametric whitening module and MoE-enhanced adaptor module, the learned universal item representations are more isotropic and suitable for domain fusion and adaptation. Especially, the results on cross-platform evaluation (\ie Online Retail dataset)  show that our approach can be effectively transferred  to a different platform via universal sequence representation pre-training. Besides, the model fine-tuned in inductive setting (without item IDs, \ourt) also has a comparable performance with other baselines. This further illustrates the effectiveness of the proposed universal SRL approach.

\begin{figure}[t]
	{
		\begin{minipage}[t]{0.99\linewidth}
			\centering
			\includegraphics[width=1\textwidth]{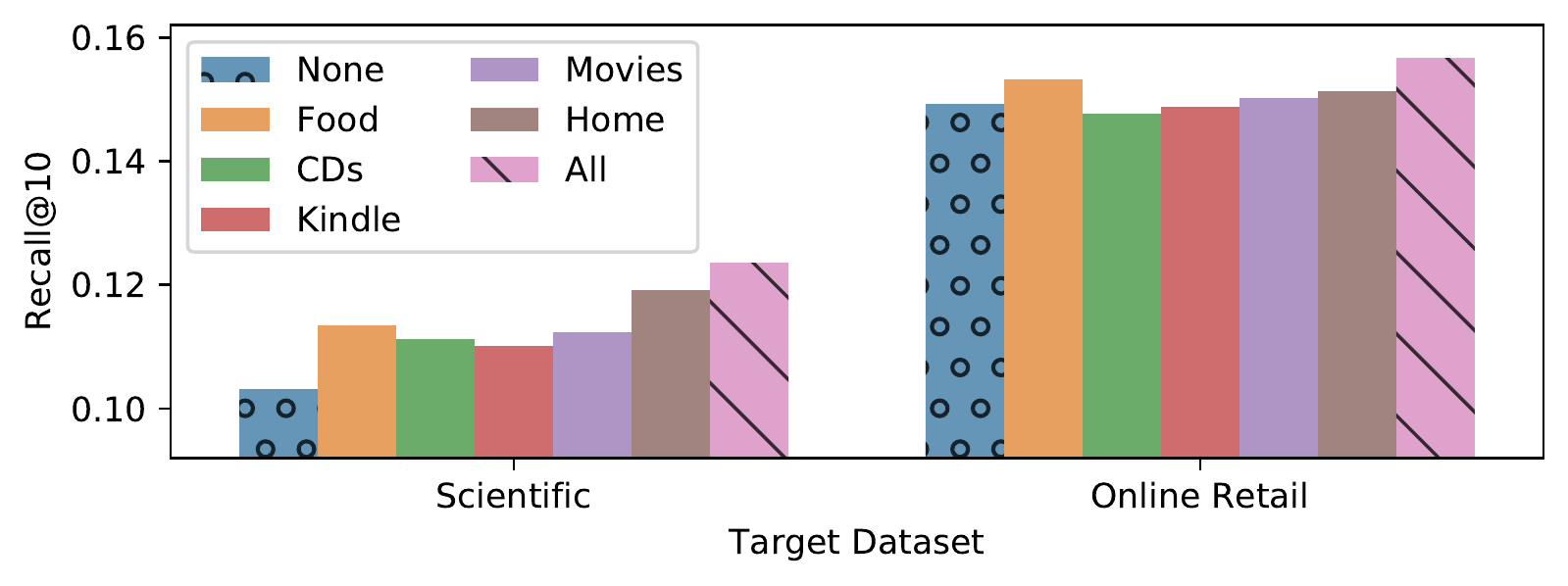}
		\end{minipage}
	}
	\caption{Performance comparison \wrt different pre-training datasets on ``Scientific'' and ``Online Retail''. ``All'' denotes the model pre-trained on all five datasets, and ``None'' denotes the training from scratch.}
	\label{fig:uni_pt}
\end{figure}

\subsection{Further Analysis}

\subsubsection{Universal Pre-training Analysis}
In this part, we compare and analyze the effectiveness of universal pre-training.
Specifically, we would like to examine whether the universal model pre-trained on multiple datasets performs better than those models pre-trained on one single source dataset, or those models without pre-training.
The experimental results are reported in Figure~\ref{fig:uni_pt}.

We can see that the model pre-trained on all the five datasets achieves better performance than any model that pre-trained on a single dataset or without pre-training. The results show that with the proposed universal sequence representation learning approach, the pre-trained model can capture semantic sequential patterns from multiple source domains to improve recommendation on target domains or platforms.

\begin{figure}[t]
	{
		\begin{minipage}[t]{0.44\linewidth}
			\centering
			\includegraphics[width=1\textwidth]{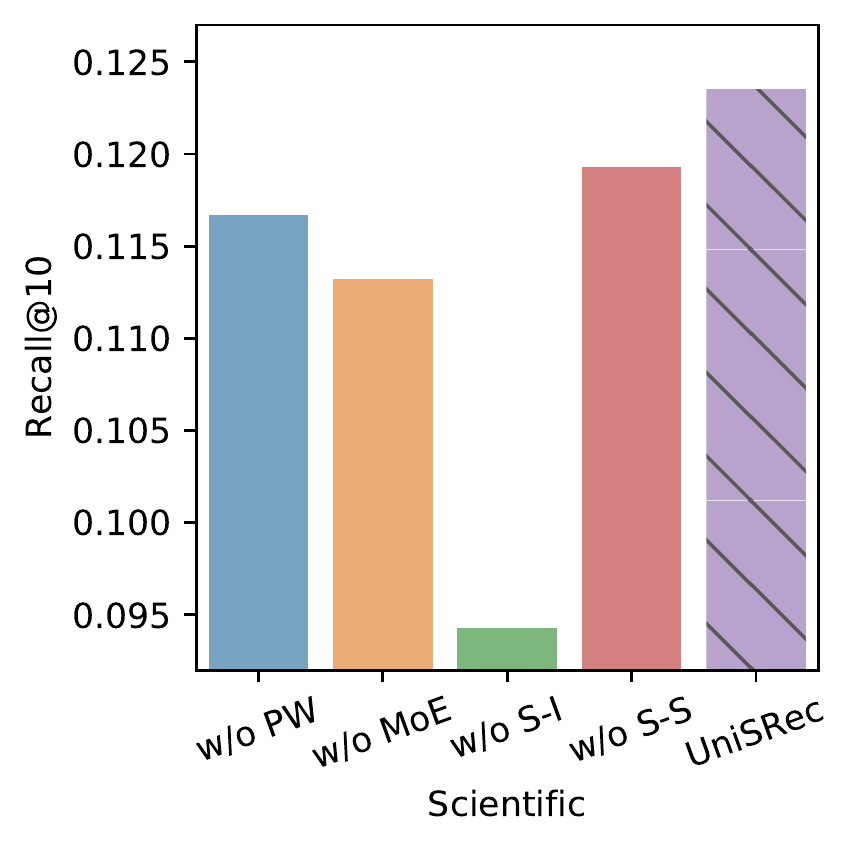}
		\end{minipage}
		\begin{minipage}[t]{0.44\linewidth}
			\centering
			\includegraphics[width=1\textwidth]{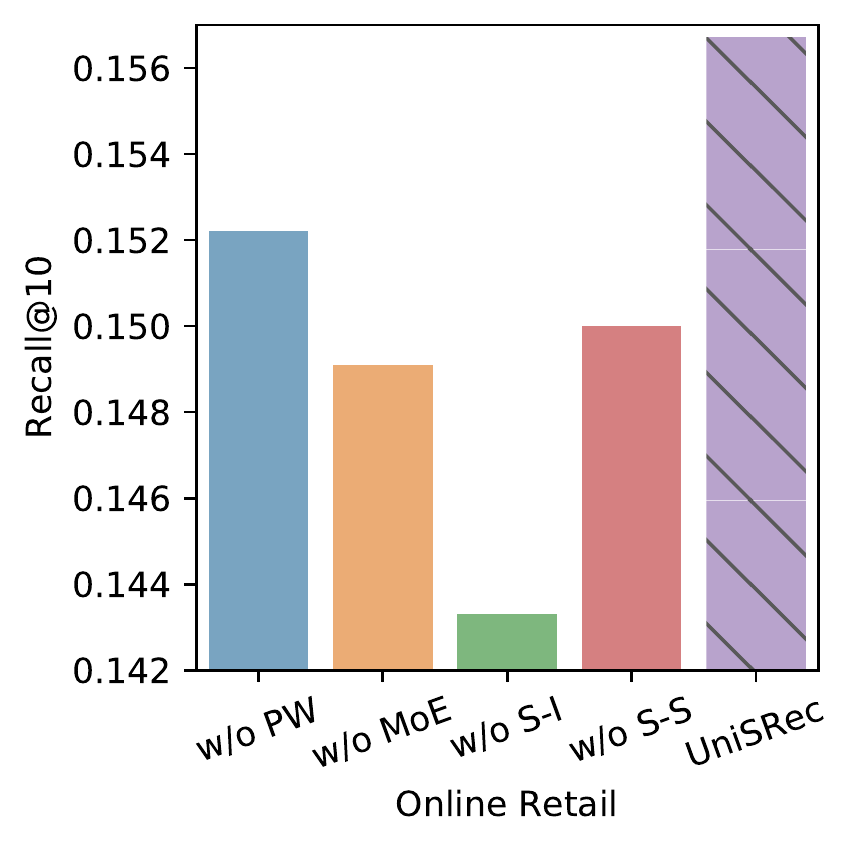}
		\end{minipage}
	}
	\caption{Ablation study of UniSRec variants on ``Scientific'' and ``Online Retail''.}
	\label{fig:ablation}
\end{figure}

\subsubsection{Ablation Study}
In this part, we analyze how each of the proposed techniques or components affects the final performance. 
We prepare four variants of the proposed UniSRec model for comparisons, including (1) \uline{$w/o$ PW} that replaces parametric whitening with traditional linear layer, \ie replacing  Eqn.~\eqref{eq:whitening} by $\widetilde{\bm{x}}_i = \bm{x}_i \cdot \bm{W}_1 -\bm{b}$, (2) \uline{$w/o$ MoE} without MoE-enhanced adaptor ($G=1$ in Eqn.~\eqref{eq:h_i}), (3)  \uline{$w/o$ S-I} without sequence-item contrastive task, and (4) \uline{$w/o$ S-S} without sequence-sequence contrastive task.

The experimental results of the proposed approach UniSRec and its variants are reported in Figure~\ref{fig:ablation}.
We can observe that all the proposed components are useful to improve the recommendation performance.
The variant $w/o$ MoE has poor performances because the MoE-enhance adaptor is the key component to improve the representation capacity for domain fusion and adaptation.

\begin{figure}[t]
	{
		\begin{minipage}[t]{0.49\linewidth}
			\centering
			\includegraphics[width=1\textwidth]{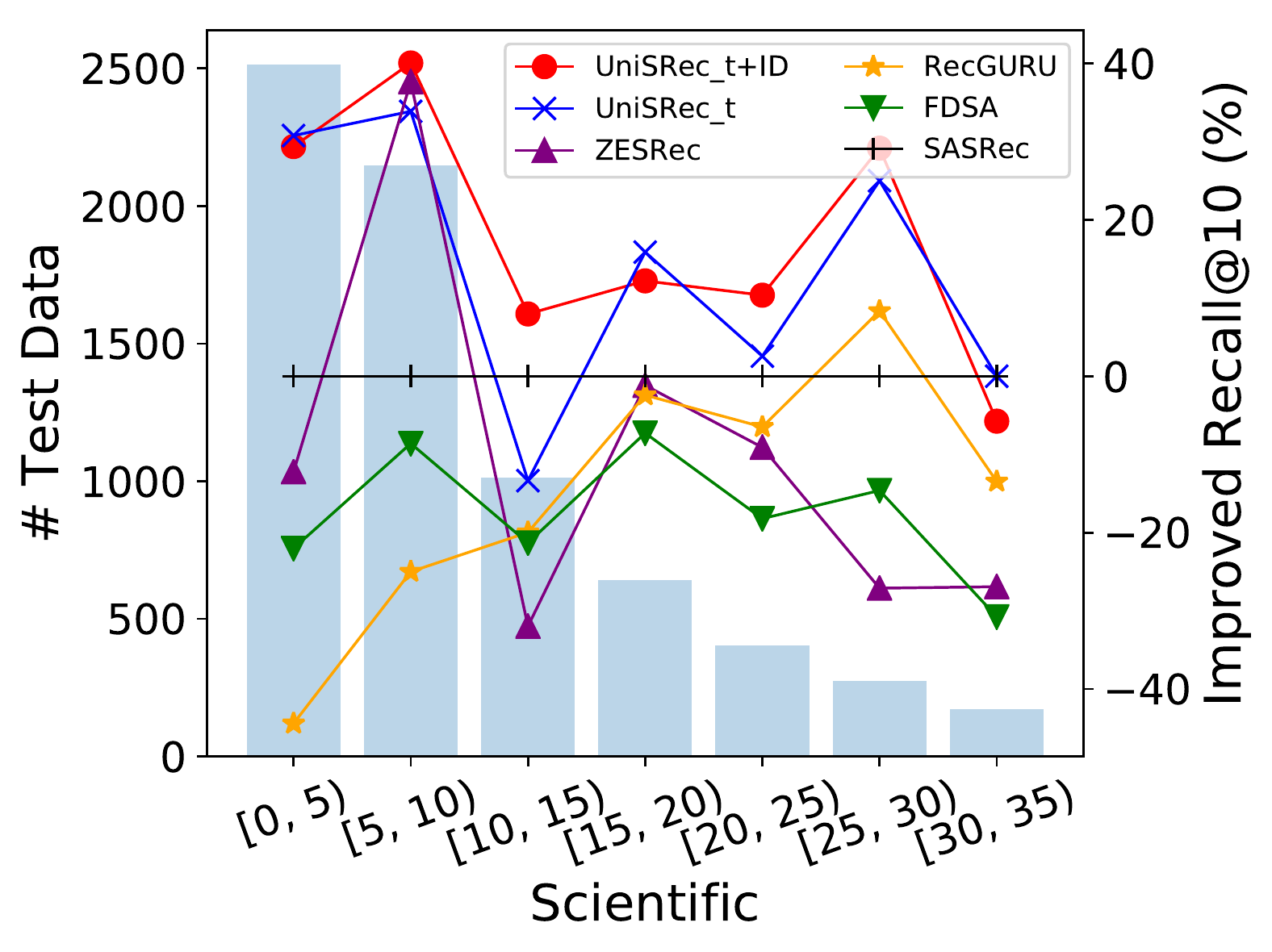}
		\end{minipage}
		\begin{minipage}[t]{0.49\linewidth}
			\centering
			\includegraphics[width=1\textwidth]{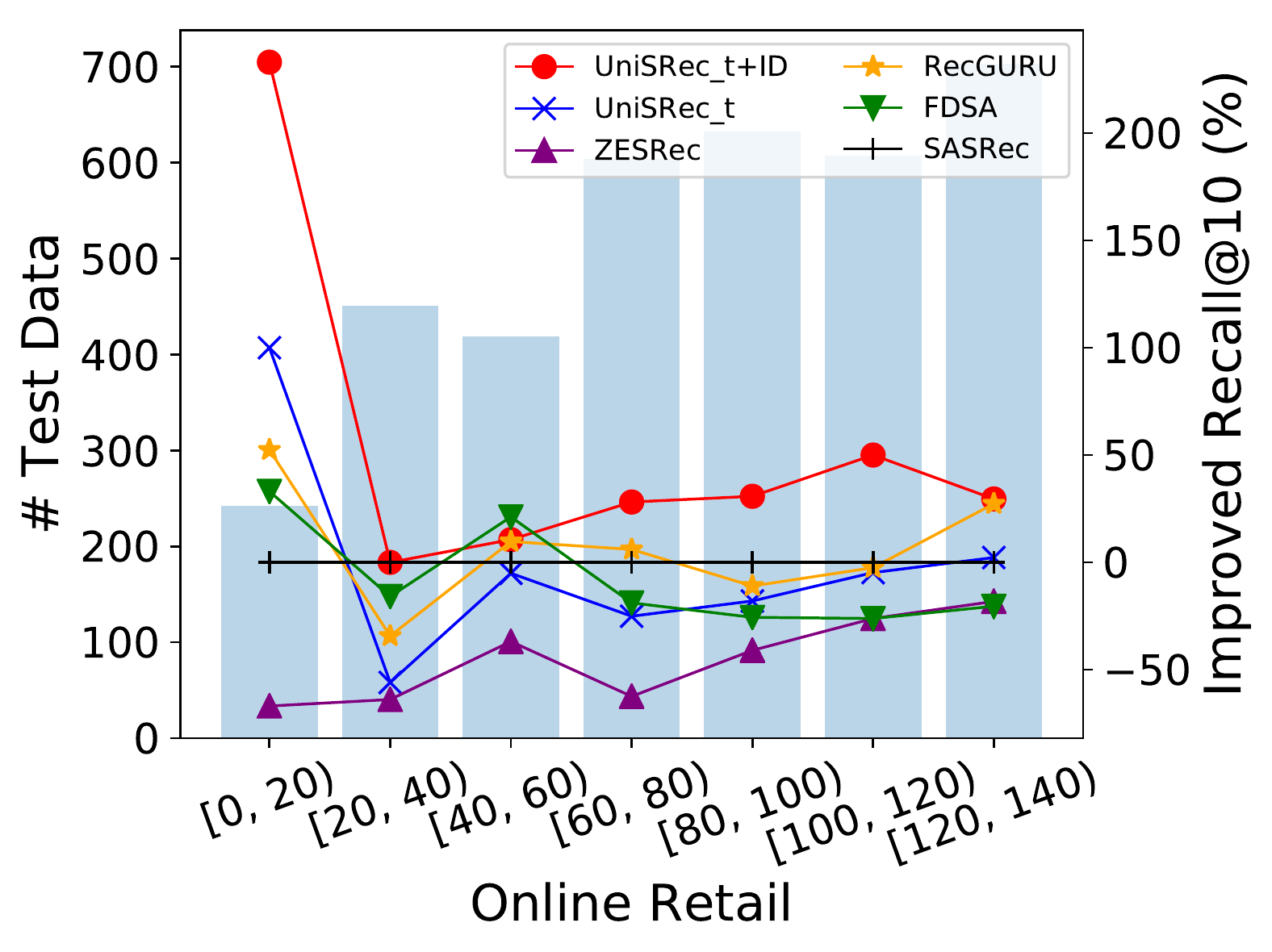}
		\end{minipage}
	}
	\caption{Performance comparison \wrt long-tail items on the ``Scientific'' and ``Online Retail'' datasets. The bar graph represents the number of interactions in test data for each group. The line chart represents the improvement ratios for Recall@10 compared with SASRec.}
	\label{fig:cold_start}
\end{figure}

\subsubsection{Performance Comparison \wrt Long-tail Items.}
One motivation to learn universal and transferable sequence representations is to alleviate the cold-start recommendation issue.
To verify this, we split the test data into different groups according to the popularity of ground-truth items in the training data, and then compare the improved ratio of Recall@10 score (\emph{w.r.t.} the baseline  SASRec) in each group.
From Figure~\ref{fig:cold_start}, we can observe that the proposed approach outperforms the other baseline models in most cases, especially when the ground truth item is unpopular, \eg group [0, 5) on Industrial and Scientific datasets and group [0, 20) on Online Retail dataset.
The results show that long-tail items can benefit from the learned universal sequence representations.

\subsection{Case Study}

\begin{figure}[t]
    \centering
    \includegraphics[width=0.39\textwidth]{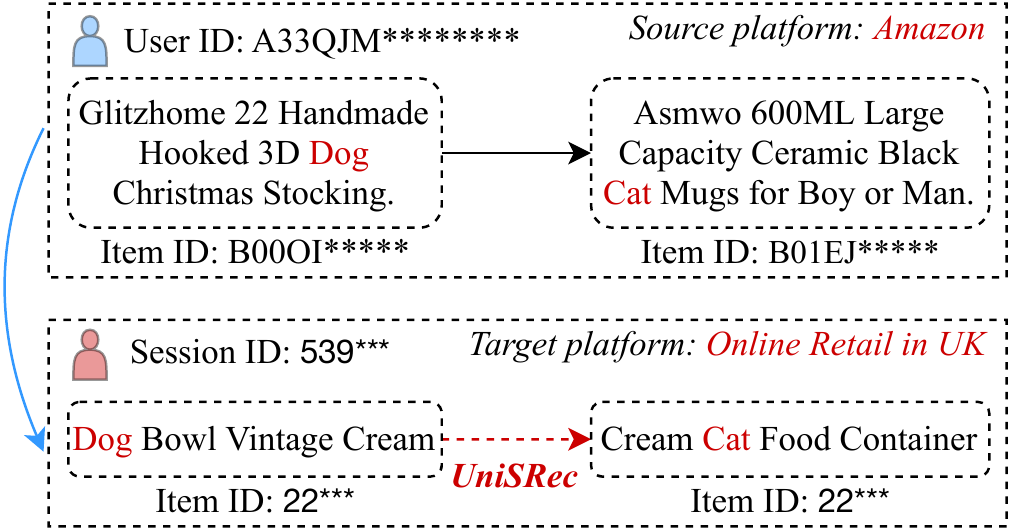}
    \caption{The purchase history of a user in the source platform (top) and the purchase history within an anonymous session in the target platform (bottom). There are naturally no overlapping users and items between the two platforms. This case shows that UniSRec can capture the universal semantic sequence pattern (\eg ``Dog $\rightarrow$ Cat'') from large-scale pre-training,
    which helps improve the recommendation performance of the target platform.}
    \label{fig:case}
\end{figure}

As shown in Table~\ref{tab:exp-main}, we can see that our approach can achieve good performance in a cross-platform setting (from \emph{Amazon} to \emph{Online Retail}). To our knowledge, in the literature of recommender systems, there are few studies that can conduct cross-platform recommendation~\cite{zhao2014weknow}. 
It is interesting to study what kind of knowledge is actually transferred across different platforms. 
For this purpose, we present an illustrative case in Figure~\ref{fig:case}. 

This example presents two short sequences of two different users from Amazon and Online Retail, respectively.
It can be observed that our approach doesn't rely on explicit item IDs to capture the user preference. Instead, it tries to capture the semantic associations by modeling the sequential patterns. Especially, the two short sequences correspond to a semantic transition from the keyword of ``\emph{dog}'' to the keyword ``\emph{cat}'' as shown in the item title. It shows that our approach can capture universal sequential patterns across platforms in terms of general textual semantics. 

Note that there might be also semantic correlations among other keywords.
Here, we select the keywords of ``{dog}'' and ``{cat}'' just for simplifying the illustration. We leave a deep investigation  of the universal representations as future work.

\section{Related Work}

\paratitle{Sequential recommendation.}
To alleviate information overload, recommender systems are widely studied and deployed in various real-world services. 
Specially, it has been shown that sequential behaviors are important signals to reflect user preferences, and thus sequential recommendation has recieved much attention from both research and industry community~\cite{rendle2010fpmc, hidasi2016gru4rec}. Early works on this topic adopt the Markov Chain assumption by estimating item-item transition probability matrices~\cite{rendle2010fpmc}. With the development of deep learning, \citet{hidasi2016gru4rec} firstly introduce Gated Recurrent Units (GRU) to model sequential behaviors. Then various kinds of neural network techniques are proposed to encode interaction sequences, such as
Recurrent Neural Network (RNN)~\cite{li2017narm},
Transformer~\cite{kang2018sasrec,sun2019bert4rec,he2021locker,hou2022core}, Multilayer Perceptron (MLP)~\cite{zhou2022fmlp}
and Graph Neural Network (GNN)~\cite{wu2019srgnn,chang2021gnn}.
Besides directly mining sequential patterns of IDs, some works are proposed to model sequences with rich features~\cite{zhang2019fdsa,zhou2020s3rec} or additional self-supervised signals~\cite{yu2022self,zhou2020s3rec,xie2022cl4rec}.
However, item representations and model parameters of these methods are usually restricted to specific data domains or platforms, making it difficult to leverage multi-domain data to improve sequential recommendation.

\paratitle{Transfer learning in recommender systems.}
To deal with the data sparsity and cold-start issues in the recommender systems, various works aim to leverage behavior information from other domains~\cite{zhu2021crossdomain} or platforms~\cite{lin2019cross} to improve recommendation performance of the target domain or both domains~\cite{zhu2019dtcdr}.
Most works rely on explicit overlapping data for conducting the transfer across domains, such as common users~\cite{hu2018conet} or items~\cite{singh2008cmf,zhu2019dtcdr}, social networks~\cite{lin2019cross} and attributes~\cite{tang2012cdcr}.
Recently, some works attempt to learn universal user representations for different user-oriented downstream tasks~\cite{yuan2020peterrec}.
However, the need of explicitly shared data still limits the application scope of the above methods.
In this work, we propose to represent items in a universal way via text-based PLMs for domain adaptation.
Different from the recently proposed zero-shot recommendation methods~\cite{ding2021zero}, we do not require that the source and target domains are closely related.
With carefully designed universal pre-training tasks and MoE-enhanced adaptor architecture, the proposed universal SRL approach can be pre-trained on data from multiple source domains, and further generalize to different domains without explicitly shared anchors.

\section{Conclusions}
In this paper, we propose the universal sequence representation learning approach for recommender systems, named \emph{UniSRec}.
Different from existing sequential recommendation methods that reply on explicit item IDs for representation learning, the proposed approach UniSRec utilizes item texts to learn  more transferable representations for sequential recommendation.
Specifically, we design a lightweight architecture based on parametric whitening and MoE-enhanced adaptor to learn the universal item representations. We further design two contrastive pre-training tasks to learn universal sequence representations from multi-domain sequences.
For future work, we will consider collecting more recommendation data to train larger user behavior models.
Besides, we will explore more kinds of  side information to represent items and improve sequence representation learning, \eg images and videos.

\begin{acks}
This work was partially supported by National Natural Science Foundation of China under Grant No. 61872369,
Beijing Natural Science Foundation under Grant No. 4222027,  and Beijing Outstanding Young Scientist Program under Grant No. BJJWZYJH012019100020098.
This work was partially supported by Beijing Academy of Artificial Intelligence~(BAAI).
Yupeng Hou was also partially supported by Alibaba Group through Alibaba Research Intern Program.
Xin Zhao is the corresponding author.
\end{acks}

\bibliographystyle{ACM-Reference-Format}
\balance
\bibliography{main}


\begin{thebibliography}{39}


\ifx \showCODEN    \undefined \def \showCODEN     #1{\unskip}     \fi
\ifx \showDOI      \undefined \def \showDOI       #1{#1}\fi
\ifx \showISBNx    \undefined \def \showISBNx     #1{\unskip}     \fi
\ifx \showISBNxiii \undefined \def \showISBNxiii  #1{\unskip}     \fi
\ifx \showISSN     \undefined \def \showISSN      #1{\unskip}     \fi
\ifx \showLCCN     \undefined \def \showLCCN      #1{\unskip}     \fi
\ifx \shownote     \undefined \def \shownote      #1{#1}          \fi
\ifx \showarticletitle \undefined \def \showarticletitle #1{#1}   \fi
\ifx \showURL      \undefined \def \showURL       {\relax}        \fi
\providecommand\bibfield[2]{#2}
\providecommand\bibinfo[2]{#2}
\providecommand\natexlab[1]{#1}
\providecommand\showeprint[2][]{arXiv:#2}

\bibitem[Bengio et~al\mbox{.}(2013)]%
        {bengio2013rl}
\bibfield{author}{\bibinfo{person}{Yoshua Bengio}, \bibinfo{person}{Aaron~C.
  Courville}, {and} \bibinfo{person}{Pascal Vincent}.}
  \bibinfo{year}{2013}\natexlab{}.
\newblock \showarticletitle{Representation Learning: {A} Review and New
  Perspectives}.
\newblock \bibinfo{journal}{\emph{{TPAMI}}} (\bibinfo{year}{2013}).
\newblock


\bibitem[Chang et~al\mbox{.}(2021)]%
        {chang2021gnn}
\bibfield{author}{\bibinfo{person}{Jianxin Chang}, \bibinfo{person}{Chen Gao},
  \bibinfo{person}{Yu Zheng}, \bibinfo{person}{Yiqun Hui},
  \bibinfo{person}{Yanan Niu}, \bibinfo{person}{Yang Song},
  \bibinfo{person}{Depeng Jin}, {and} \bibinfo{person}{Yong Li}.}
  \bibinfo{year}{2021}\natexlab{}.
\newblock \showarticletitle{Sequential Recommendation with Graph Neural
  Networks}. In \bibinfo{booktitle}{\emph{{SIGIR}}}.
\newblock


\bibitem[Devlin et~al\mbox{.}(2019)]%
        {devlin2019bert}
\bibfield{author}{\bibinfo{person}{Jacob Devlin}, \bibinfo{person}{Ming-Wei
  Chang}, \bibinfo{person}{Kenton Lee}, {and} \bibinfo{person}{Kristina
  Toutanova}.} \bibinfo{year}{2019}\natexlab{}.
\newblock \showarticletitle{BERT: Pre-training of Deep Bidirectional
  Transformers for Language Understanding}. In
  \bibinfo{booktitle}{\emph{{NAACL}}}.
\newblock


\bibitem[Ding et~al\mbox{.}(2021)]%
        {ding2021zero}
\bibfield{author}{\bibinfo{person}{Hao Ding}, \bibinfo{person}{Yifei Ma},
  \bibinfo{person}{Anoop Deoras}, \bibinfo{person}{Yuyang Wang}, {and}
  \bibinfo{person}{Hao Wang}.} \bibinfo{year}{2021}\natexlab{}.
\newblock \showarticletitle{Zero-Shot Recommender Systems}.
\newblock \bibinfo{journal}{\emph{arXiv preprint arXiv:2105.08318}}
  (\bibinfo{year}{2021}).
\newblock


\bibitem[He et~al\mbox{.}(2021)]%
        {he2021locker}
\bibfield{author}{\bibinfo{person}{Zhankui He}, \bibinfo{person}{Handong Zhao},
  \bibinfo{person}{Zhe Lin}, \bibinfo{person}{Zhaowen Wang},
  \bibinfo{person}{Ajinkya Kale}, {and} \bibinfo{person}{Julian McAuley}.}
  \bibinfo{year}{2021}\natexlab{}.
\newblock \showarticletitle{Locally constrained self-attentive sequential
  recommendation}. In \bibinfo{booktitle}{\emph{{CIKM}}}.
\newblock


\bibitem[Hidasi et~al\mbox{.}(2016)]%
        {hidasi2016gru4rec}
\bibfield{author}{\bibinfo{person}{Bal{\'{a}}zs Hidasi},
  \bibinfo{person}{Alexandros Karatzoglou}, \bibinfo{person}{Linas Baltrunas},
  {and} \bibinfo{person}{Domonkos Tikk}.} \bibinfo{year}{2016}\natexlab{}.
\newblock \showarticletitle{Session-based Recommendations with Recurrent Neural
  Networks}. In \bibinfo{booktitle}{\emph{{ICLR}}}.
\newblock


\bibitem[Hou et~al\mbox{.}(2022)]%
        {hou2022core}
\bibfield{author}{\bibinfo{person}{Yupeng Hou}, \bibinfo{person}{Binbin Hu},
  \bibinfo{person}{Zhiqiang Zhang}, {and} \bibinfo{person}{Wayne~Xin Zhao}.}
  \bibinfo{year}{2022}\natexlab{}.
\newblock \showarticletitle{CORE: Simple and Effective Session-based
  Recommendation within Consistent Representation Space}. In
  \bibinfo{booktitle}{\emph{{SIGIR}}}.
\newblock


\bibitem[Hu et~al\mbox{.}(2018)]%
        {hu2018conet}
\bibfield{author}{\bibinfo{person}{Guangneng Hu}, \bibinfo{person}{Yu Zhang},
  {and} \bibinfo{person}{Qiang Yang}.} \bibinfo{year}{2018}\natexlab{}.
\newblock \showarticletitle{CoNet: Collaborative Cross Networks for
  Cross-Domain Recommendation}. In \bibinfo{booktitle}{\emph{{CIKM}}}.
\newblock


\bibitem[Huang et~al\mbox{.}(2021)]%
        {huang2021white2}
\bibfield{author}{\bibinfo{person}{Junjie Huang}, \bibinfo{person}{Duyu Tang},
  \bibinfo{person}{Wanjun Zhong}, \bibinfo{person}{Shuai Lu},
  \bibinfo{person}{Linjun Shou}, \bibinfo{person}{Ming Gong},
  \bibinfo{person}{Daxin Jiang}, {and} \bibinfo{person}{Nan Duan}.}
  \bibinfo{year}{2021}\natexlab{}.
\newblock \showarticletitle{WhiteningBERT: An Easy Unsupervised Sentence
  Embedding Approach}. In \bibinfo{booktitle}{\emph{Findings of {EMNLP}}}.
\newblock


\bibitem[Kang and McAuley(2018)]%
        {kang2018sasrec}
\bibfield{author}{\bibinfo{person}{Wang{-}Cheng Kang} {and}
  \bibinfo{person}{Julian~J. McAuley}.} \bibinfo{year}{2018}\natexlab{}.
\newblock \showarticletitle{Self-Attentive Sequential Recommendation}. In
  \bibinfo{booktitle}{\emph{{ICDM}}}.
\newblock


\bibitem[Li et~al\mbox{.}(2020)]%
        {li2020bertflow}
\bibfield{author}{\bibinfo{person}{Bohan Li}, \bibinfo{person}{Hao Zhou},
  \bibinfo{person}{Junxian He}, \bibinfo{person}{Mingxuan Wang},
  \bibinfo{person}{Yiming Yang}, {and} \bibinfo{person}{Lei Li}.}
  \bibinfo{year}{2020}\natexlab{}.
\newblock \showarticletitle{On the Sentence Embeddings from Pre-trained
  Language Models}. In \bibinfo{booktitle}{\emph{{EMNLP}}}.
\newblock


\bibitem[Li et~al\mbox{.}(2022)]%
        {li2022recguru}
\bibfield{author}{\bibinfo{person}{Chenglin Li}, \bibinfo{person}{Mingjun
  Zhao}, \bibinfo{person}{Huanming Zhang}, \bibinfo{person}{Chenyun Yu},
  \bibinfo{person}{Lei Cheng}, \bibinfo{person}{Guoqiang Shu},
  \bibinfo{person}{Beibei Kong}, {and} \bibinfo{person}{Di Niu}.}
  \bibinfo{year}{2022}\natexlab{}.
\newblock \showarticletitle{{RecGURU}: Adversarial Learning of Generalized User
  Representations for Cross-Domain Recommendation}. In
  \bibinfo{booktitle}{\emph{{WSDM}}}.
\newblock


\bibitem[Li et~al\mbox{.}(2017)]%
        {li2017narm}
\bibfield{author}{\bibinfo{person}{Jing Li}, \bibinfo{person}{Pengjie Ren},
  \bibinfo{person}{Zhumin Chen}, \bibinfo{person}{Zhaochun Ren},
  \bibinfo{person}{Tao Lian}, {and} \bibinfo{person}{Jun Ma}.}
  \bibinfo{year}{2017}\natexlab{}.
\newblock \showarticletitle{Neural Attentive Session-based Recommendation}. In
  \bibinfo{booktitle}{\emph{{CIKM}}}.
\newblock


\bibitem[Lin et~al\mbox{.}(2019)]%
        {lin2019cross}
\bibfield{author}{\bibinfo{person}{Tzu{-}Heng Lin}, \bibinfo{person}{Chen Gao},
  {and} \bibinfo{person}{Yong Li}.} \bibinfo{year}{2019}\natexlab{}.
\newblock \showarticletitle{{CROSS:} Cross-platform Recommendation for Social
  E-Commerce}. In \bibinfo{booktitle}{\emph{{SIGIR}}}.
\newblock


\bibitem[Lin et~al\mbox{.}(2022)]%
        {lin2022ncl}
\bibfield{author}{\bibinfo{person}{Zihan Lin}, \bibinfo{person}{Changxin Tian},
  \bibinfo{person}{Yupeng Hou}, {and} \bibinfo{person}{Wayne~Xin Zhao}.}
  \bibinfo{year}{2022}\natexlab{}.
\newblock \showarticletitle{Improving Graph Collaborative Filtering with
  Neighborhood-enriched Contrastive Learning}. In
  \bibinfo{booktitle}{\emph{{TheWebConf}}}.
\newblock


\bibitem[Mu et~al\mbox{.}(2022)]%
        {mu2022ida}
\bibfield{author}{\bibinfo{person}{Shanlei Mu}, \bibinfo{person}{Yupeng Hou},
  \bibinfo{person}{Wayne~Xin Zhao}, \bibinfo{person}{Yaliang Li}, {and}
  \bibinfo{person}{Bolin Ding}.} \bibinfo{year}{2022}\natexlab{}.
\newblock \showarticletitle{ID-Agnostic User Behavior Pre-training for
  Sequential Recommendation}.
\newblock \bibinfo{journal}{\emph{arXiv preprint arXiv:2206.02323}}
  (\bibinfo{year}{2022}).
\newblock


\bibitem[Ni et~al\mbox{.}(2019)]%
        {ni2019amazon}
\bibfield{author}{\bibinfo{person}{Jianmo Ni}, \bibinfo{person}{Jiacheng Li},
  {and} \bibinfo{person}{Julian~J. McAuley}.} \bibinfo{year}{2019}\natexlab{}.
\newblock \showarticletitle{Justifying Recommendations using Distantly-Labeled
  Reviews and Fine-Grained Aspects}. In \bibinfo{booktitle}{\emph{{EMNLP}}}.
\newblock


\bibitem[Rendle et~al\mbox{.}(2010)]%
        {rendle2010fpmc}
\bibfield{author}{\bibinfo{person}{Steffen Rendle}, \bibinfo{person}{Christoph
  Freudenthaler}, {and} \bibinfo{person}{Lars Schmidt{-}Thieme}.}
  \bibinfo{year}{2010}\natexlab{}.
\newblock \showarticletitle{Factorizing personalized Markov chains for
  next-basket recommendation}. In \bibinfo{booktitle}{\emph{{WWW}}}.
\newblock


\bibitem[Shazeer et~al\mbox{.}(2017)]%
        {shazeer2017moe}
\bibfield{author}{\bibinfo{person}{Noam Shazeer}, \bibinfo{person}{Azalia
  Mirhoseini}, \bibinfo{person}{Krzysztof Maziarz}, \bibinfo{person}{Andy
  Davis}, \bibinfo{person}{Quoc~V. Le}, \bibinfo{person}{Geoffrey~E. Hinton},
  {and} \bibinfo{person}{Jeff Dean}.} \bibinfo{year}{2017}\natexlab{}.
\newblock \showarticletitle{Outrageously Large Neural Networks: The
  Sparsely-Gated Mixture-of-Experts Layer}. In
  \bibinfo{booktitle}{\emph{{ICLR}}}.
\newblock


\bibitem[Singh and Gordon(2008)]%
        {singh2008cmf}
\bibfield{author}{\bibinfo{person}{Ajit~Paul Singh} {and}
  \bibinfo{person}{Geoffrey~J. Gordon}.} \bibinfo{year}{2008}\natexlab{}.
\newblock \showarticletitle{Relational learning via collective matrix
  factorization}. In \bibinfo{booktitle}{\emph{{SIGKDD}}}.
\newblock


\bibitem[Su et~al\mbox{.}(2021)]%
        {su2021whitening}
\bibfield{author}{\bibinfo{person}{Jianlin Su}, \bibinfo{person}{Jiarun Cao},
  \bibinfo{person}{Weijie Liu}, {and} \bibinfo{person}{Yangyiwen Ou}.}
  \bibinfo{year}{2021}\natexlab{}.
\newblock \showarticletitle{Whitening Sentence Representations for Better
  Semantics and Faster Retrieval}.
\newblock \bibinfo{journal}{\emph{arXiv preprint arXiv:2103.15316}}
  (\bibinfo{year}{2021}).
\newblock


\bibitem[Sun et~al\mbox{.}(2019)]%
        {sun2019bert4rec}
\bibfield{author}{\bibinfo{person}{Fei Sun}, \bibinfo{person}{Jun Liu},
  \bibinfo{person}{Jian Wu}, \bibinfo{person}{Changhua Pei},
  \bibinfo{person}{Xiao Lin}, \bibinfo{person}{Wenwu Ou}, {and}
  \bibinfo{person}{Peng Jiang}.} \bibinfo{year}{2019}\natexlab{}.
\newblock \showarticletitle{BERT4Rec: Sequential Recommendation with
  Bidirectional Encoder Representations from Transformer}. In
  \bibinfo{booktitle}{\emph{{CIKM}}}.
\newblock


\bibitem[Tang et~al\mbox{.}(2020)]%
        {tang2020ple}
\bibfield{author}{\bibinfo{person}{Hongyan Tang}, \bibinfo{person}{Junning
  Liu}, \bibinfo{person}{Ming Zhao}, {and} \bibinfo{person}{Xudong Gong}.}
  \bibinfo{year}{2020}\natexlab{}.
\newblock \showarticletitle{Progressive Layered Extraction {(PLE):} {A} Novel
  Multi-Task Learning {(MTL)} Model for Personalized Recommendations}. In
  \bibinfo{booktitle}{\emph{{RecSys}}}.
\newblock


\bibitem[Tang and Wang(2018)]%
        {tang2018caser}
\bibfield{author}{\bibinfo{person}{Jiaxi Tang} {and} \bibinfo{person}{Ke
  Wang}.} \bibinfo{year}{2018}\natexlab{}.
\newblock \showarticletitle{Personalized Top-N Sequential Recommendation via
  Convolutional Sequence Embedding}. In \bibinfo{booktitle}{\emph{{WSDM}}}.
\newblock


\bibitem[Tang et~al\mbox{.}(2012)]%
        {tang2012cdcr}
\bibfield{author}{\bibinfo{person}{Jie Tang}, \bibinfo{person}{Sen Wu},
  \bibinfo{person}{Jimeng Sun}, {and} \bibinfo{person}{Hang Su}.}
  \bibinfo{year}{2012}\natexlab{}.
\newblock \showarticletitle{Cross-domain collaboration recommendation}. In
  \bibinfo{booktitle}{\emph{{SIGKDD}}}.
\newblock


\bibitem[Vaswani et~al\mbox{.}(2017)]%
        {vaswani2017attention}
\bibfield{author}{\bibinfo{person}{Ashish Vaswani}, \bibinfo{person}{Noam
  Shazeer}, \bibinfo{person}{Niki Parmar}, \bibinfo{person}{Jakob Uszkoreit},
  \bibinfo{person}{Llion Jones}, \bibinfo{person}{Aidan~N. Gomez},
  \bibinfo{person}{Lukasz Kaiser}, {and} \bibinfo{person}{Illia Polosukhin}.}
  \bibinfo{year}{2017}\natexlab{}.
\newblock \showarticletitle{Attention is All you Need}. In
  \bibinfo{booktitle}{\emph{{NeurIPS}}}.
\newblock


\bibitem[Wu et~al\mbox{.}(2019)]%
        {wu2019srgnn}
\bibfield{author}{\bibinfo{person}{Shu Wu}, \bibinfo{person}{Yuyuan Tang},
  \bibinfo{person}{Yanqiao Zhu}, \bibinfo{person}{Liang Wang},
  \bibinfo{person}{Xing Xie}, {and} \bibinfo{person}{Tieniu Tan}.}
  \bibinfo{year}{2019}\natexlab{}.
\newblock \showarticletitle{Session-Based Recommendation with Graph Neural
  Networks}. In \bibinfo{booktitle}{\emph{{AAAI}}}.
\newblock


\bibitem[Xie et~al\mbox{.}(2022a)]%
        {xie2021ccdr}
\bibfield{author}{\bibinfo{person}{Ruobing Xie}, \bibinfo{person}{Qi Liu},
  \bibinfo{person}{Liangdong Wang}, \bibinfo{person}{Shukai Liu},
  \bibinfo{person}{Bo Zhang}, {and} \bibinfo{person}{Leyu Lin}.}
  \bibinfo{year}{2022}\natexlab{a}.
\newblock \showarticletitle{Contrastive Cross-domain Recommendation in
  Matching}. In \bibinfo{booktitle}{\emph{{SIGKDD}}}.
\newblock


\bibitem[Xie et~al\mbox{.}(2022b)]%
        {xie2022cl4rec}
\bibfield{author}{\bibinfo{person}{Xu Xie}, \bibinfo{person}{Fei Sun},
  \bibinfo{person}{Zhaoyang Liu}, \bibinfo{person}{Shiwen Wu},
  \bibinfo{person}{Jinyang Gao}, \bibinfo{person}{Bolin Ding}, {and}
  \bibinfo{person}{Bin Cui}.} \bibinfo{year}{2022}\natexlab{b}.
\newblock \showarticletitle{Contrastive learning for sequential
  recommendation}. In \bibinfo{booktitle}{\emph{{ICDE}}}.
\newblock


\bibitem[Yu et~al\mbox{.}(2022)]%
        {yu2022self}
\bibfield{author}{\bibinfo{person}{Junliang Yu}, \bibinfo{person}{Hongzhi Yin},
  \bibinfo{person}{Xin Xia}, \bibinfo{person}{Tong Chen},
  \bibinfo{person}{Jundong Li}, {and} \bibinfo{person}{Zi Huang}.}
  \bibinfo{year}{2022}\natexlab{}.
\newblock \showarticletitle{Self-Supervised Learning for Recommender Systems: A
  Survey}.
\newblock \bibinfo{journal}{\emph{arXiv preprint arXiv:2203.15876}}
  (\bibinfo{year}{2022}).
\newblock


\bibitem[Yuan et~al\mbox{.}(2020)]%
        {yuan2020peterrec}
\bibfield{author}{\bibinfo{person}{Fajie Yuan}, \bibinfo{person}{Xiangnan He},
  \bibinfo{person}{Alexandros Karatzoglou}, {and} \bibinfo{person}{Liguang
  Zhang}.} \bibinfo{year}{2020}\natexlab{}.
\newblock \showarticletitle{Parameter-Efficient Transfer from Sequential
  Behaviors for User Modeling and Recommendation}. In
  \bibinfo{booktitle}{\emph{{SIGIR}}}.
\newblock


\bibitem[Zhang et~al\mbox{.}(2019)]%
        {zhang2019fdsa}
\bibfield{author}{\bibinfo{person}{Tingting Zhang}, \bibinfo{person}{Pengpeng
  Zhao}, \bibinfo{person}{Yanchi Liu}, \bibinfo{person}{Victor~S. Sheng},
  \bibinfo{person}{Jiajie Xu}, \bibinfo{person}{Deqing Wang},
  \bibinfo{person}{Guanfeng Liu}, {and} \bibinfo{person}{Xiaofang Zhou}.}
  \bibinfo{year}{2019}\natexlab{}.
\newblock \showarticletitle{Feature-level Deeper Self-Attention Network for
  Sequential Recommendation}. In \bibinfo{booktitle}{\emph{{IJCAI}}}.
\newblock


\bibitem[Zhao et~al\mbox{.}(2020)]%
        {zhao2020revisiting}
\bibfield{author}{\bibinfo{person}{Wayne~Xin Zhao}, \bibinfo{person}{Junhua
  Chen}, \bibinfo{person}{Pengfei Wang}, \bibinfo{person}{Qi Gu}, {and}
  \bibinfo{person}{Ji-Rong Wen}.} \bibinfo{year}{2020}\natexlab{}.
\newblock \showarticletitle{Revisiting Alternative Experimental Settings for
  Evaluating Top-N Item Recommendation Algorithms}. In
  \bibinfo{booktitle}{\emph{{CIKM}}}.
\newblock


\bibitem[Zhao et~al\mbox{.}(2014)]%
        {zhao2014weknow}
\bibfield{author}{\bibinfo{person}{Wayne~Xin Zhao}, \bibinfo{person}{Yanwei
  Guo}, \bibinfo{person}{Yulan He}, \bibinfo{person}{Han Jiang},
  \bibinfo{person}{Yuexin Wu}, {and} \bibinfo{person}{Xiaoming Li}.}
  \bibinfo{year}{2014}\natexlab{}.
\newblock \showarticletitle{We know what you want to buy: a demographic-based
  system for product recommendation on microblogs}. In
  \bibinfo{booktitle}{\emph{{KDD}}}.
\newblock


\bibitem[Zhao et~al\mbox{.}(2021)]%
        {zhao2021recbole}
\bibfield{author}{\bibinfo{person}{Wayne~Xin Zhao}, \bibinfo{person}{Shanlei
  Mu}, \bibinfo{person}{Yupeng Hou}, \bibinfo{person}{Zihan Lin},
  \bibinfo{person}{Yushuo Chen}, \bibinfo{person}{Xingyu Pan},
  \bibinfo{person}{Kaiyuan Li}, \bibinfo{person}{Yujie Lu},
  \bibinfo{person}{Hui Wang}, \bibinfo{person}{Changxin Tian},
  \bibinfo{person}{Yingqian Min}, \bibinfo{person}{Zhichao Feng},
  \bibinfo{person}{Xinyan Fan}, \bibinfo{person}{Xu Chen},
  \bibinfo{person}{Pengfei Wang}, \bibinfo{person}{Wendi Ji},
  \bibinfo{person}{Yaliang Li}, \bibinfo{person}{Xiaoling Wang}, {and}
  \bibinfo{person}{Ji-Rong Wen}.} \bibinfo{year}{2021}\natexlab{}.
\newblock \showarticletitle{RecBole: Towards a Unified, Comprehensive and
  Efficient Framework for Recommendation Algorithms}. In
  \bibinfo{booktitle}{\emph{{CIKM}}}.
\newblock


\bibitem[Zhou et~al\mbox{.}(2020)]%
        {zhou2020s3rec}
\bibfield{author}{\bibinfo{person}{Kun Zhou}, \bibinfo{person}{Hui Wang},
  \bibinfo{person}{Wayne~Xin Zhao}, \bibinfo{person}{Yutao Zhu},
  \bibinfo{person}{Sirui Wang}, \bibinfo{person}{Fuzheng Zhang},
  \bibinfo{person}{Zhongyuan Wang}, {and} \bibinfo{person}{Ji{-}Rong Wen}.}
  \bibinfo{year}{2020}\natexlab{}.
\newblock \showarticletitle{S3-Rec: Self-Supervised Learning for Sequential
  Recommendation with Mutual Information Maximization}. In
  \bibinfo{booktitle}{\emph{{CIKM}}}.
\newblock


\bibitem[Zhou et~al\mbox{.}(2022)]%
        {zhou2022fmlp}
\bibfield{author}{\bibinfo{person}{Kun Zhou}, \bibinfo{person}{Hui Yu},
  \bibinfo{person}{Wayne~Xin Zhao}, {and} \bibinfo{person}{Ji{-}Rong Wen}.}
  \bibinfo{year}{2022}\natexlab{}.
\newblock \showarticletitle{Filter-enhanced {MLP} is All You Need for
  Sequential Recommendation}. In \bibinfo{booktitle}{\emph{{TheWebConf}}}.
\newblock


\bibitem[Zhu et~al\mbox{.}(2019)]%
        {zhu2019dtcdr}
\bibfield{author}{\bibinfo{person}{Feng Zhu}, \bibinfo{person}{Chaochao Chen},
  \bibinfo{person}{Yan Wang}, \bibinfo{person}{Guanfeng Liu}, {and}
  \bibinfo{person}{Xiaolin Zheng}.} \bibinfo{year}{2019}\natexlab{}.
\newblock \showarticletitle{{DTCDR:} {A} Framework for Dual-Target Cross-Domain
  Recommendation}. In \bibinfo{booktitle}{\emph{{CIKM}}}.
\newblock


\bibitem[Zhu et~al\mbox{.}(2021)]%
        {zhu2021crossdomain}
\bibfield{author}{\bibinfo{person}{Feng Zhu}, \bibinfo{person}{Yan Wang},
  \bibinfo{person}{Chaochao Chen}, \bibinfo{person}{Jun Zhou},
  \bibinfo{person}{Longfei Li}, {and} \bibinfo{person}{Guanfeng Liu}.}
  \bibinfo{year}{2021}\natexlab{}.
\newblock \showarticletitle{Cross-Domain Recommendation: Challenges, Progress,
  and Prospects}. In \bibinfo{booktitle}{\emph{{IJCAI}}}.
\newblock


\end{thebibliography}

\end{document}